\documentclass[11pt, prd, aps, amssymb, amsmath, tightenlines, twocolumn]{revtex4-2}

\usepackage[utf8]{inputenc}
\usepackage{graphicx}
\usepackage{xcolor}
\usepackage{amsmath}
\usepackage{amsfonts}
\usepackage{amssymb}
\usepackage{adjustbox}
\usepackage{multirow}
\usepackage{lineno}
\usepackage{tikz}
\usepackage{textcomp}
\usepackage{tabularx}
\usepackage{rotating}
\usepackage{fancyhdr}
\usepackage{booktabs}
\usepackage{array}

\begin{document}

\pagestyle{plain}

\title{Energy shift and decay width of kaonium from coupled-channel one-loop chiral amplitudes}

\author{Alireza Beygi}\email{alirezabeygi389@gmail.com}
\affiliation{Independent Researcher, Ederstr. 14, 60486 Frankfurt am Main, Germany}

\date{\today}

\begin{abstract}
Using two fitted coupled-channel one-loop chiral amplitudes, we investigate the strong-interaction energy shift and decay width of the hypothetical Coulomb-bound $K^+ K^-$ hadronic atom, kaonium. We obtain ground-state widths of 1.67 and 0.92 keV, which are both larger than previous kaonium predictions, including the 0.30 keV result obtained from a non-perturbative treatment based on leading-order chiral perturbation theory. Although the $K^+ K^-$ scattering lengths obtained from the two one-loop amplitudes have nearly equal imaginary parts, their real parts have opposite signs; thus, the leading absorptive contributions to the widths are similar, while the Coulomb corrections act in opposite directions. The kaonium width therefore depends on both the real and imaginary parts of the near-threshold strong amplitude. Despite the increased ground-state width, the first three $s$-wave levels remain separated. In photon fusion, at fixed pole area and resonance position, the increase in the intrinsic width has a negligible effect on the line shape after including Belle's experimental resolution.
\end{abstract}

\maketitle

\section{Introduction}
\label{In}

The study of the exotic Coulomb-bound $K^+ K^-$ mesonic atom, kaonium, which is yet to be observed experimentally, provides a useful way to probe the near-threshold $K \bar{K}$ interaction and the dynamics associated with the $f_0 ( 980 )$ and $a_0 ( 980 )$ resonances. These resonances strongly influence the near-threshold $K \bar{K}$ amplitudes and hence the $K^+ K^-$ scattering length required to calculate the strong-interaction energy shift and decay width of kaonium. The binding of this hadronic atom is mainly Coulombic, while the strong interaction shifts the Coulomb energy levels and induces their decay. The resulting complex atomic energy therefore depends on both the real and imaginary parts of the near-threshold $K \bar{K}$ amplitudes. Because the charged and neutral kaons have different masses, the $K^0 \bar{K}^0$ threshold lies 7.9 MeV above the $K^+ K^-$ threshold. Although the charge exchange channel $K^+ K^- \to K^0 \bar{K}^0$ is kinematically closed at the charged threshold, its virtual coupling modifies the physical $K^+ K^-$ scattering length. This effect plays a crucial role in obtaining the kaonium spectrum.

The properties of kaonium have previously been studied using various approaches; see \cite{SW1993, BK1995, SB1996, SK2004, YZ2006, RHL2009, SPK2011}. For example, vector meson exchange interactions are used in \cite{SK2004}, and they are extended to include momentum dependence in \cite{RHL2009}; a non-perturbative treatment based on leading-order (LO) chiral perturbation theory is discussed in \cite{SPK2011}. The reported ground-state decay widths in these papers are substantially different, but all remain below 1 keV. For example, the LO chiral calculation of \cite{SPK2011} gives a width of about 0.30 keV, while \cite{SK2004} reports values of about 0.21 and 0.58 keV. More recently, kaonium formation has been studied in the photon-fusion reactions $\gamma \gamma \to \pi^0 \pi^0$ and $\gamma \gamma \to \pi^0 \eta$ \cite{AB2026}.

In the present work, we determine the strong-interaction energy shift and decay width of kaonium using two fitted coupled-channel one-loop chiral amplitudes unitarized with the inverse amplitude method (IAM) \cite{AGN2002}. We use Fit 2 of \cite{YJS2021} and the earlier IAM fit of \cite{AGN2002}, later referred to as IAM I \cite{JRP2004}, as two one-loop descriptions of the near-threshold $K \bar{K}$ interaction. We obtain ground-state widths of 1.67 and 0.92 keV for Fit 2 and IAM I, respectively, both larger than previous kaonium predictions. We show that the difference between these widths stems from the Coulomb corrections associated with the two charged scattering lengths.

To find out whether the low-lying levels still remain spectrally distinct, we calculate the first three $s$-wave levels. We also study the electromagnetic $2 p \to 1 s$ transition and re-examine kaonium production in photon fusion. We show that at fixed pole area and resonance position, the increased width has a negligible effect on the line shape once Belle's experimental resolution is taken into account.

The paper is organized as follows. In Sec.~\ref{II}, we describe the two coupled-channel one-loop chiral amplitudes used as the strong-interaction input and determine the physical $K^+ K^-$ threshold parameters. In Sec.~\ref{III}, we introduce the Coulomb pole equation, the strong-Coulomb matching procedure, and the leading electromagnetic corrections. In Sec.~\ref{IV}, we present the ground-state energy shifts and decay widths, and examine the origin of the difference between the two one-loop results. Therein, we also investigate the excited-state spectrum and transitions, and the Fit 2 parameter dependence and systematic variations. We distinguish the pole width from the real-axis full width at half maximum (FWHM) in Sec.~\ref{V} and examine the photon-fusion line shape, and discuss the kaonium lifetime and propagation in matter. We conclude in Sec.~\ref{VI}. Numerical details are given in Appendix~\ref{app}.

\section{Strong-interaction input}
\label{II}

\subsection{Coupled-channel one-loop formalism}
\label{IIA}

We use two previously fitted coupled-channel one-loop chiral amplitudes as the strong-interaction input for the kaonium calculation. Both are based on the complete renormalized one-loop SU(3) chiral amplitudes and use the coupled-channel inverse amplitude method (IAM) \cite{AGN2002}. For both amplitudes, we adopt the scalar partial-wave normalization of \cite{AGN2002},
\begin{equation}
t^{I}_{a b} (s) = \frac{1}{32 N \pi} \int_{-1}^{1} dx \, T^{I}_{a b} \bigl( s, t(s, x), u(s, x) \bigr).
\label{pwn}
\end{equation}
Here $T^{I}_{a b} (s, t, u)$ is the invariant amplitude for the process $a \to b$ with isospin $I$, while $t^{I}_{a b} (s)$ denotes its scalar partial wave; $s, t, u$ are the Mandelstam variables, and $x = \cos \theta$, with $\theta$ the center-of-mass scattering angle. The factor $N = 2$ corresponds to the identical $\pi \pi$ and $\eta \eta$ elastic channels and $N = 1$ otherwise. We order the channels as $(K \bar{K}, \pi \pi, \eta \eta)$ for $I = 0$, and $(K \bar{K}, \pi \eta)$ for $I = 1$. Writing the chiral expansion as $t = t_2 + t_4 + O( p^6 )$, perturbative unitarity on the physical cut gives $\operatorname{Im} t_4 = t_2 \Sigma t_2^\ast, \quad \Sigma_{a b} = \delta_{a b} \sigma_a \theta ( s - s_a^{\rm th} )$, where $\sigma_a$ is the corresponding two-body phase space factor and $s_a^{\rm th}$ is the threshold of channel $a$. The coupled-channel IAM amplitude is $t_{\rm IAM} = t_2 (t_2 - t_4)^{-1} t_2$. Expanding $t_{\rm IAM}$ reproduces $t_2 + t_4$ through $O( p^4 )$. The resummed amplitude satisfies coupled-channel unitarity, $\operatorname{Im} t_{\rm IAM}^{-1} = -\Sigma$, when the perturbative unitarity relation above holds in the channel space considered.

\subsection{Fit 2 and IAM I}
\label{IIB}

For the main calculation, we use Fit 2 of \cite{YJS2021}. In the scalar sectors relevant here, it includes the coupled $\pi \pi$, $K \bar{K}$, and $\eta \eta$ channels for $I = 0$ and the $\pi \eta$ and $K \bar{K}$ channels for $I = 1$. We use the masses, pion decay constant, renormalization scale, fitted low-energy constants, and prescription of \cite{YJS2021} when evaluating the threshold amplitudes. In that prescription, the imaginary parts of the relevant $t$- and $u$-channel loop contributions are removed in order to restore coupled-channel unitarity below the highest threshold. The details of the global fit are given in \cite{YJS2021}.

For comparison, we use the one-loop IAM fit of \cite{AGN2002}, referred to as IAM I following the terminology of \cite{JRP2004}. IAM I assumes $f_\pi$ in the amplitudes, while IAM II uses $f_K$ and $f_\eta$ where appropriate, together with the corresponding next-to-leading-order terms; IAM III is a variant of IAM II. The amplitudes used here are those of IAM I. Note that IAM I gives an overly narrow $f_0 (980)$ and no nearby $a_0 (980)$ pole \cite{JRP2004}; we use it here, as mentioned, for the purpose of comparing with Fit 2.

\subsection{IAM I pole and Riemann sheet}
\label{IIC}

As a check of the IAM I implementation, we adopt the following continuation prescription for the $I = 0$ amplitude. Between the $\pi \pi$ and $K \bar{K}$ thresholds, only the $\pi \pi$ channel is open. With the matrix channel order $( K \bar{K}, \pi \pi, \eta \eta )$, we write
\begin{equation}
t^{-1}_{(-, +, +)}(s) = \left.t^{-1}_{\rm I}(s) \right|_{\rm lower} - 2 i \, \operatorname{diag} \left( 0, \sigma_{\pi \pi}(s), 0 \right),
\label{IAMPC}
\end{equation}
where $t_{\mathrm I}$ denotes the IAM I amplitude evaluated with the principal branches of the loop functions, and lower corresponds to its evaluation in the lower half-plane. The sheet signs are ordered as $( \pi \pi, K \bar{K}, \eta \eta )$. Because of overlapping left-hand cuts, we use sheet II as a label for the continuation prescription in (\ref{IAMPC}); see also \cite{AGN2002, YJS2021}. The pole is obtained by solving $\det t^{-1}_{(-, +, +)}(s) = 0$ in the complex $s$ plane, giving
\begin{equation}
\sqrt{s_{f_0}} = (988.07 - 3.45 \, i) \ {\rm MeV},
\label{focu}
\end{equation}
in agreement with the published rounded IAM I value \cite{JRP2004},
\begin{equation}
\sqrt{s_{f_0}} \simeq (988 - 4 \, i) \ {\rm MeV}.
\label{pri}
\end{equation}

\subsection{Threshold parameters and neutral kaon cusp}
\label{IID}

With the partial-wave normalization defined in Sec.~\ref{IIA}, the corresponding non-relativistic $s$-wave amplitude for equal kaon masses is $f_I = t^I_{K \bar{K}} / E_K$, where $E_K = \sqrt{m_K^2 + k^2}$. We determine the complex scattering length and effective range from the effective range expansion \cite{HAB1949},
\begin{equation}
\frac{\sqrt{1 + q^2}}{t^I_{K \bar{K}}(s)} + i q = -\frac{1}{\bar{a}_I} + \frac{\bar{r}_I}{2} q^2 + O( q^4 ),
\label{ERE}
\end{equation}
where $s = 4 m_K^2 ( 1 + q^2 )$, $q = k / m_K$, $\bar{a}_I = m_K a_I$, and $\bar{r}_I = m_K r_I$. Our scattering length convention is $f_I ( 0 ) = -a_I$. The scattering lengths and effective ranges are complex because lighter meson channels are open at the $K \bar{K}$ threshold.

Using the kaon mass convention of each fit, the isospin scattering lengths obtained from the two one-loop amplitudes are:
\begin{equation}
\begin{array}{c|cc}
& \bar{a}_0 & \bar{a}_1 \\
\hline
{\rm IAM \ I} & 9.276 - 3.893 \, i & -1.099 - 1.917 \, i \\
{\rm Fit \ 2} & 4.905 - 13.590 \, i & -1.100 - 1.513 \, i
\end{array}.
\label{isoscatt}
\end{equation}
These are isospin-limit quantities. The nearby $K^0 \bar{K}^0$ threshold modifies the physical $K^+ K^-$ scattering length entering the atomic calculation.

At the $K^+ K^-$ threshold, the $K^0 \bar{K}^0$ channel is closed. Eliminating the closed neutral channel gives \cite{SPK2011, AB2026},
\begin{equation}
a_p = \frac{a - \kappa_0 a_0 a_1}{1 - \kappa_0 a},
\label{apcusp}
\end{equation}
where $a = ( a_0 + a_1 ) / 2$ and $\kappa_0 = \sqrt{m_{K^0}^2 - m_{K^+}^2}$. The relation for $a_p$ includes the leading neutral kaon cusp associated with the kaon mass difference. Using the physical masses $m_{K^+} = 493.677$ MeV and $m_{K^0} = 497.611$ MeV \cite{PDG2026}, which give $\kappa_0 = 62.448$ MeV, we obtain for IAM I and Fit 2, respectively,
\begin{equation}
\begin{aligned}
\bar{a}_p &= 7.072 - 7.809 \, i \quad ( {\rm IAM \ I} ), \\
\bar{a}_p &= -2.790 - 7.645 \, i \quad ( {\rm Fit \ 2} ),
\end{aligned}
\label{chargedlengths}
\end{equation}
where $\bar{a}_p = m_{K^+} a_p$. The imaginary parts of the expressions in (\ref{chargedlengths}) differ by only about $2.1\%$, whereas their real parts have opposite signs. As shown in Sec.~\ref{IV}, the opposite signs of the real parts lead to Coulomb corrections of opposite sign in the two kaonium widths.

The effective range in the physical charged channel is determined by keeping the $O( k^2 )$ terms when eliminating the neutral channel, using the same effective range convention as above. For the Fit 2 best fit, we obtain
\begin{equation}
\bar{r}_p = -3.973 + 0.958 \, i,
\label{fit2range}
\end{equation}
where $\bar{r}_p = m_{K^+} r_p$. The corresponding effective range correction to the atomic result is small and is treated separately from the central scattering length result.

We treat the shift from the isospin-limit threshold to the physical charged kaon threshold as a separate systematic variation. Defining
\begin{equation}
q_+^2 = \frac{m_{K^+}^2}{m_K^2} - 1,
\end{equation}
we re-expand the inverse scattering length as
\begin{equation}
\frac{1}{\bar{a}_I^{(+)}} = \frac{1}{\bar{a}_I} - \frac{\bar{r}_I}{2} q_+^2.
\label{thresholdreexp}
\end{equation}
For Fit 2, $q_+^2 = -7.927 \times 10^{-3}$. We then apply the cusp relation to the re-expanded scattering lengths. Direct evaluation of the inverse amplitude on the physical rim of the cut at the charged threshold gives atomic quantities that agree with the re-expansion within 0.3 eV.

\section{Coulomb pole and electromagnetic corrections}
\label{III}

\subsection{Coulomb pole equation}
\label{IIIA}

We next include the attractive Coulomb interaction. We define the dimensionless matching radius $\bar{d} = m_{K^+} d$. Following \cite{SPK2011}, we take $\bar{d} = 2.2$, corresponding to $d = 0.879$ fm, as the reference value.

The strong and Coulomb-modified scattering lengths are related by \cite{XK2000, SPK2011},
\begin{equation}
\frac{1}{\bar{a}_p^C} = \frac{1}{\bar{a}_p} - \alpha \left[ \ln( \alpha \bar{d} ) + \gamma_E \right],
\label{xks}
\end{equation}
where $\bar{a}_p^C = m_{K^+} a_p^C$ and $\gamma_E$ is Euler's constant. For the reduced mass $\mu_c = m_{K^+} / 2$, we introduce $\lambda ( k ) = - i k / ( \mu_c \alpha )$ and $k = i \mu_c \alpha \lambda$ at a bound-state pole. This fixes the outgoing-wave convention used below.

Using the effective range convention explained in Sec.~\ref{IID}, the charged channel amplitude satisfies
\begin{equation}
-\left[ \bar{f}_p^{-1}(q) + i q \right] = \frac{1}{\bar{a}_p} - \frac{\bar{r}_p}{2} q^2 + O( q^4 ),
\label{erca}
\end{equation}
where $\bar{f}_p = m_{K^+} f_p$ and $q = k / m_{K^+}$. At a bound-state pole, $q = i \alpha \lambda / 2$. The corresponding Kudryavtsev--Popov (KP) eigenvalue equation can then be written as \cite{AEK1979, AEK19792, SK2004, SPK2011},
\begin{equation}
\begin{aligned}
&D_{\rm KP} (\lambda) = \frac{1}{\bar{a}_p^C} + \frac{\bar{r}_p \alpha^2 \lambda^2}{8} \\
&-\alpha \left[ \frac{\lambda}{2} + \ln \lambda + \psi \left( 1 - \frac{1}{\lambda} \right) + \gamma_E \right] = 0.
\end{aligned}
\label{kp1}
\end{equation}
We fix the logarithm branch by continuity from the Coulomb limit and track each KP root from $\lambda = 1 / n$ as the strong interaction is turned on.

For a solution $\lambda_n$, the complex atomic energy is $E_n = -\mu_c \alpha^2 \lambda_n^2 / 2$, while the point-Coulomb energy is $E_n^C = -\mu_c \alpha^2 / ( 2 n^2 )$. We define the complex strong-interaction energy shift by
\begin{equation}
\delta E_n \equiv E_n - E_n^C = \Delta E_n^{\rm str} - \frac{i}{2} \Gamma_n,
\label{dele1}
\end{equation}
so that
\begin{equation}
\Delta E_n^{\rm str} = {\rm Re} \, ( E_n - E_n^C ), \quad \Gamma_n = -2 \, {\rm Im} \, E_n.
\label{stde}
\end{equation}
For the ground state, $E_{1 s}^C = -6572.24$ eV.

The dependence of the width on the real part of the scattering length can be seen directly by expanding the KP equation about the Coulomb 1$s$ pole. Neglecting the effective range, this gives a Trueman-type expansion \cite{TLT1961}, whose leading term is the Deser relation \cite{SD1954}:
\begin{equation}
\begin{aligned}
\frac{\delta E_{1 s}}{E_{1 s}^C} = -2 \alpha \bar{a}_p &\left[ 1 - \alpha \bar{a}_p \left( 1 - \gamma_E - \ln( \alpha \bar{d} ) \right) \right] \\
&+O \! \left[ ( \alpha \bar{a}_p )^3 \right].
\end{aligned}
\label{tde}
\end{equation}
At $\bar{d} = 2.2$, the first two terms can be written as
\begin{equation}
\begin{aligned}
{\delta E}_{1 s} &= C_1 \bar{a}_p + C_2 \bar{a}_p^2 + \cdots, \\
( C_1, C_2 ) &= ( 95.920, -3.1880 ) \ {\rm eV}.
\end{aligned}
\label{exn1}
\end{equation}
Writing $\bar{a}_p = x + i y$, we obtain
\begin{equation}
\begin{aligned}
\Delta E_{1 s}^{\rm str} &= 95.920 x - 3.1880 ( x^2 - y^2 ) + \cdots, \\
\Gamma_{1 s} &= -191.840 y + 12.752 x y + \cdots.
\end{aligned}
\label{expn2}
\end{equation}
For an absorptive amplitude, $y < 0$. The quadratic term therefore decreases the width when $x > 0$ and increases it when $x < 0$. Since IAM I has ${\rm Re} \, \bar{a}_p > 0$, whereas Fit 2 has ${\rm Re} \, \bar{a}_p < 0$, (\ref{expn2}) already shows why the Coulomb correction suppresses the IAM I width but enhances the Fit 2 width. The numerical results below are obtained by solving the full KP equation rather than from the truncated expansion; effective range effects are treated separately.

At this order, the finite Coulomb correction depends on the convention used to define the threshold amplitude. In the non-relativistic effective field theory convention of \cite{UGM2004, JG2008}, the corresponding logarithmic coefficient is
\begin{equation}
L_{\rm NREFT} = 1 - \ln \alpha = 5.9202,
\label{nrefte}
\end{equation}
whereas the present KP prescription gives
\begin{equation}
L_{\rm KP} = 1 - \gamma_E - \ln ( \alpha \bar{d} ) = 4.5546,
\end{equation}
at $\bar{d} = 2.2$. Matching the two expansions through this order gives
\begin{equation}
\bar{A} = \bar{a}_d + \alpha ( \gamma_E + \ln \bar{d} ) \bar{a}_d^2 + O ( \alpha^2 ),
\label{madt}
\end{equation}
where $\bar{a}_d$ denotes the length entering the KP boundary condition and $\bar{A}$ the corresponding dimensionless threshold amplitude in the NREFT convention. For our central results, we set $\bar{a}_d = \bar{a}_p$ at $\bar{d} = 2.2$, thereby also fixing the finite Coulomb term, which cannot be obtained from the strong scattering length alone. For comparison, we also consider $\bar{A} = \bar{a}_p$. We implement this alternative matching prescription by setting $1 / \bar{a}_d = 1 / \bar{a}_p + \alpha ( \gamma_E + \ln \bar{d} )$ in the full KP calculation. This prescription agrees with (\ref{madt}) through first order in $\alpha$. The effects of this alternative matching prescription and of varying $\bar{d}$ are discussed in Sec.~\ref{IVD}.

\subsection{Electromagnetic corrections}
\label{IIIB}

The strong-interaction shift defined in the previous subsection is measured relative to the point-Coulomb level. Here we consider the leading electromagnetic corrections beyond the point-Coulomb interaction. For the ground state, where $a_B = ( \mu_c \alpha )^{-1} = 109.55 \ {\rm fm}$ and $\xi = m_e a_B = 0.2837$, the leading correction is due to electronic vacuum polarization. In first-order perturbation theory, the Uehling potential \cite{EAU1935} gives
\begin{equation}
\begin{aligned}
\delta E_{1 s}^{\rm VP} &= -\frac{2 \alpha^2}{3 \pi a_B} \\
&\times \int_1^\infty dt \frac{\sqrt{t^2 - 1}}{t^2} \left( 1 + \frac{1}{2 t^2} \right) \frac{1}{( 1 + \xi t)^2},
\end{aligned}
\label{UVP}
\end{equation}
which yields
\begin{equation}
\delta E_{1 s}^{\rm VP} = -13.31 \ {\rm eV}.
\label{vp1s}
\end{equation}
The general vacuum polarization formula for the energy shift \cite{DE2000}, evaluated for kaonium, gives the same result.

A second correction arises from the finite charge radius of the kaon. Expanding the charged kaon electromagnetic form factors about zero momentum transfer gives the leading finite-size contribution \cite{JLF1979},
\begin{equation}
\delta E_{1 s}^{\rm fs} = \frac{2 \pi \alpha}{3} |\psi_{1 s}(0)|^2 \left( \langle r_{K^+}^2 \rangle + \langle r_{K^-}^2 \rangle \right),
\label{smc}
\end{equation}
where $|\psi_{1 s}(0)|^2 = (\pi a_B^3)^{-1}$. Using $r_{K^\pm} = 0.560(31)$ fm \cite{PDG2026}, we obtain
\begin{equation}
\delta E_{1 s}^{\rm fs} = +0.46(5) \ {\rm eV}.
\label{wb1s}
\end{equation}

Therefore, the leading additive electromagnetic correction to the point-Coulomb level is
\begin{equation}
\delta E_{1 s}^{\rm VP} + \delta E_{1 s}^{\rm fs} = -12.85(5) \ {\rm eV}.
\label{tcf}
\end{equation}
This contribution should be added to $\Delta E_{1 s}^{\rm str}$ when quoting the total level displacement relative to the point-Coulomb energy.

We conclude this section by noting that vacuum polarization also modifies the wave function at the origin, which affects the short-range strong-interaction contribution to the shift and width. For $K^+ K^-$, \cite{DE2000} gives $\delta \psi(0) / \alpha \psi(0) = 0.5158$, corresponding to an increase of 0.753\% in $|\psi(0)|^2$. We estimate the corresponding change in the strong-interaction shift and width by applying this change to the leading Deser contribution and, separately, to the full KP result. For Fit 2, this gives, respectively,
\begin{equation}
\begin{aligned}
( \delta \Delta E_{1 s}^{\rm str}, \delta \Gamma_{1 s} ) &= (-2.0, +11.0) \ {\rm eV}, \\
( \delta \Delta E_{1 s}^{\rm str}, \delta \Gamma_{1 s} ) &= (-0.5, +12.6) \ {\rm eV}.
\end{aligned}
\label{fI2}
\end{equation}
We include these estimates separately among the systematic variations in Sec.~\ref{IVD}.

\section{Results}
\label{IV}

\subsection{Ground-state shift and width}
\label{IVA}

We have summarized the ground-state kaonium results for the IAM I and Fit 2 amplitudes in Table~\ref{tab1}, together with the LO result of \cite{SPK2011} for comparison. The published LO charged scattering length already includes the neutral kaon cusp. Reference \cite{SPK2011} quotes $m_K a_p = 2.688 - 1.896 i$ for $m_K = 496$ MeV; expressing the same dimensionful scattering length in the present $m_{K^+}$ normalization gives $\bar{a}_p = 2.675 - 1.887 i$. For IAM I and Fit 2, we use the charged scattering lengths obtained in Sec.~\ref{IID}.
\begin{table}[t]
\caption{Ground-state kaonium results for the three strong-interaction inputs. The atomic quantities are obtained from the full KP equation at $\bar{d} = 2.2$. Here $B_{1 s} = -{\rm Re} \, E_{1 s}$ and $\tau_{1 s} = \hbar /\Gamma_{1 s}$. The quoted energy shifts are the strong-interaction shifts; the additive electromagnetic correction discussed in Sec.~\ref{IIIB} is not included.}
\centering \scriptsize
\begin{tabular}{@{}lccc@{}}
\toprule
Quantity & LO & IAM I & Fit 2 \\
\midrule
$\bar{a}_0$ & $3.530 - 2.077 \, i$ & $9.276 - 3.893 \, i$ & $4.905 - 13.590 \, i$ \\
$\bar{a}_1$ & $1.651 - 1.416 \, i$ & $-1.099 - 1.917 \, i$ & $-1.100 - 1.513 \, i$ \\
$\bar{a}_p$ & $2.675 - 1.887 \, i$ & $7.072 - 7.809 \, i$ & $-2.790 - 7.645 \, i$ \\
$B_{1 s}$ [keV] & 6.3278 & 5.9205 & 6.6395 \\
$\Delta E^{\rm str}_{1 s}$ [eV] & 244.5 & 651.7 & $-67.3$ \\
$\Gamma_{1 s}$ [eV] & 303.3 & 920.2 & 1673.4 \\
$\tau_{1 s}$ [$10^{-18}$ s] & 2.17 & 0.715 & 0.393 \\
\bottomrule
\end{tabular}
\label{tab1}
\end{table}

The IAM I and Fit 2 amplitudes give ground-state widths of 0.92 and 1.67 keV, respectively, both larger than the LO value of 0.30 keV. The origin of the difference between the two one-loop widths is examined in detail in the following subsection.

\subsection{Origin of the width difference}
\label{IVB}

\begin{figure}[t]
\centering
\includegraphics[width = 0.49\textwidth]{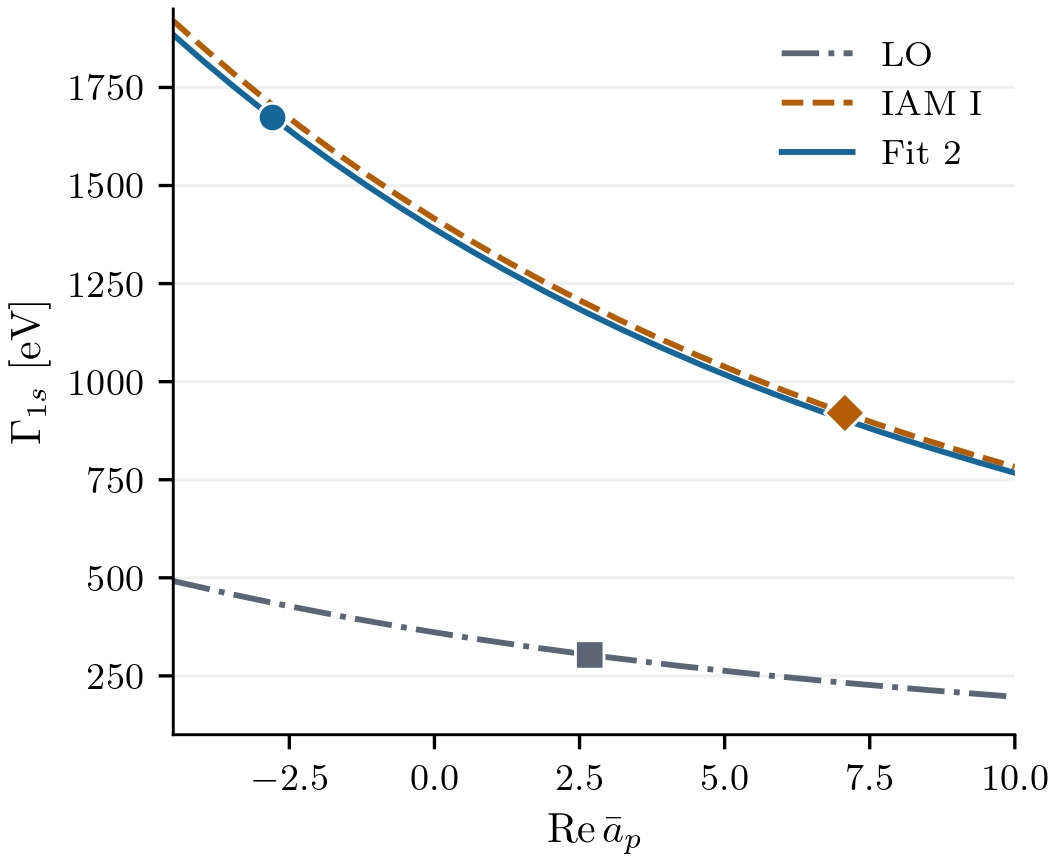}
\caption{Dependence of the full KP width on ${\rm Re} \, \bar{a}_p$, varied at fixed ${\rm Im} \, \bar{a}_p$ for each strong-interaction input. The markers indicate the corresponding LO, IAM I, and Fit 2 values. The IAM I and Fit 2 curves nearly coincide because their imaginary parts are similar.}
\label{fig1}
\end{figure}

Table~\ref{tab2} separates the leading, quadratic, and higher-order contributions to the ground-state strong shift and pole width.
\begin{table}[t]
\caption{Contributions to the 1$s$ strong shift and pole width, in eV, at $\bar{d} = 2.2$. Higher denotes the full KP result minus the leading and quadratic terms.}
\centering \small
\begin{tabular}{@{}llrrrr@{}}
\toprule
Input & Quantity & Leading & Quadratic & Higher & Full \\
\midrule
LO & $\Delta E^{\rm str}$ & 256.6 & $-11.5$ & $-0.7$ & 244.5 \\
&$\Gamma$ & 362.0 & $-64.4$ & 5.7 & 303.3 \\
IAM I & $\Delta E^{\rm str}$ & 678.3 & 35.0 & $-61.5$ & 651.7 \\
&$\Gamma$ & 1498.1 & $-704.2$ & 126.4 & 920.2 \\
Fit 2 & $\Delta E^{\rm str}$ & $-267.6$ & 161.5 & 38.8 & $-67.3$ \\
&$\Gamma$ & 1466.6 & 272.0 & $-65.1$ & 1673.4 \\
\bottomrule
\end{tabular}
\label{tab2}
\end{table}
For IAM I and Fit 2, the leading Deser widths are 1498.1 and 1466.6 eV, respectively, and are therefore nearly equal. However, the quadratic terms change the widths by $-704.2$ and $+272.0$ eV, and the higher-order KP contributions by $+126.4$ and $-65.1$ eV. Hence, the Coulomb corrections change two similar leading absorptive widths into substantially different full widths. As follows from (\ref{expn2}), the sign of ${\rm Re} \, \bar{a}_p$ determines the direction of the quadratic correction: it suppresses the IAM I width and enhances the Fit 2 width. Figure~\ref{fig1} illustrates this dependence directly.

The same decomposition also explains the small Fit 2 strong shift, $-267.6 + 161.5 + 38.8 = -67.3 \ {\rm eV}$. The final shift is much smaller than the individual contributions and is particularly sensitive to variations of the strong input.

We next quantify the sensitivity of the width to the underlying isospin amplitudes. At the Fit 2 point, its local dependence is
\begin{equation}
\begin{aligned}
\frac{\delta \Gamma_{1 s}}{{\rm eV}} \simeq &80.4 \, \delta{\rm Re} \, \bar{a}_0 - 60.4 \, \delta{\rm Im} \, \bar{a}_0 \\
&-230.3 \, \delta{\rm Re} \, \bar{a}_1 - 29.3 \, \delta{\rm Im} \, \bar{a}_1.
\end{aligned}
\label{feis}
\end{equation}
A derivation is given in Appendix~\ref{app}. Each coefficient corresponds to varying one component while keeping the other three fixed. The width is most sensitive to $\operatorname{Re} \bar{a}_1$, showing that the isovector amplitude is also important for the kaonium width.
For comparison, varying the charged scattering length directly at fixed ${\rm Im} \, \bar{a}_p$ gives
\begin{equation}
\frac{\partial \Gamma_{1 s}} {\partial{\rm Re} \, \bar{a}_p} = -114.5 \ {\rm eV}.
\label{ffeis}
\end{equation}
Thus, the width is controlled not only by the absorptive part of the charged amplitude but also by its real part through the Coulomb correction.

\subsection{Excited states and transitions}
\label{IVC}

We also examine the 2$s$ and 3$s$ poles of the same KP equation. For Fit 2, in the scattering length approximation, the first three $s$-wave levels are summarized in Table~\ref{tabiii}.
\begin{table}[t]
\caption{First three Fit 2 $s$-wave poles at $\bar{d} = 2.2$, in the scattering length approximation. The energies are measured relative to the charged kaon threshold, while the strong shifts are measured relative to the corresponding point-Coulomb levels.}
\centering \small
\begin{tabular}{@{}lrrr@{}}
\toprule
Level & $\Delta E_n^{\rm str}$ [eV] & $\Gamma_n$ [eV] & $\operatorname{Re} E_n$ [keV] \\
\midrule
1$s$ & $-67.3$ & 1673.4 & $-6.6395$ \\
2$s$ & $-13.7$ & 206.9 & $-1.6567$ \\
3$s$ & $-4.5$ & 61.0 & $-0.7348$ \\
\bottomrule
\end{tabular}
\label{tabiii}
\end{table}
The width ratios are
\begin{equation}
\frac{\Gamma_{1 s}}{\Gamma_{2 s}} = 8.09, \qquad \frac{\Gamma_{1 s}}{\Gamma_{3 s}} = 27.43,
\label{sec1e}
\end{equation}
within 2\% of the leading Coulomb scaling $| \psi_{n s} (0) |^2 \propto n^{-3}$, which gives the ratios 8 and 27. The corresponding shift ratios are 4.93 and 14.83, showing a larger departure from $1 / n^3$ scaling. This reflects the additional $n$ dependence of the Coulomb correction and the cancellation that reduces the Fit 2 ground-state shift.

For Fit 2,
\begin{equation}
\frac{\Gamma_{1 s}}{B_{1 s}} = 0.252, \qquad \frac{\Gamma_{1 s}} {{\rm Re} \, E_{2 s} - {\rm Re} \, E_{1 s}} = 0.336.
\label{sec2e}
\end{equation}
The latter ratio is 0.064 for LO and 0.211 for IAM I, showing that the Fit 2 ground state is broader relative to the level spacing. But still, the first three $s$-wave levels remain separated and become narrower with increasing principal quantum number. The ground-state pole also remains below threshold, with $B_{1 s} / \Gamma_{1 s} = 3.97$, and corresponds to a short-lived quasi-bound state.

A further measure of the separation of neighboring unstable levels is obtained by comparing their energy spacing with the sum of their widths. For the 1$s$ and 2$s$ levels, the separation is 4.983 keV, and the summed width is 1.880 keV, corresponding to 0.377 of the separation. For the 2$s$ and 3$s$ levels, the corresponding values are 0.922 and 0.268 keV. These ratios provide another indication that the low-lying $s$-wave levels remain distinct despite the larger Fit 2 width.

Electric dipole transitions between two $s$-wave states are forbidden. For higher partial waves, the strong level shift is governed by the corresponding threshold parameter. We define the complex $p$-wave scattering volume $v_p$ by
\begin{equation}
k^3 \cot \delta_1 = -\frac{1}{v_p} + \cdots,
\label{sec3e}
\end{equation}
and then evaluate the higher partial-wave pseudo-potential of \cite{ZI2006} in a Coulomb $n p$ state, which gives, to leading order,
\begin{equation}
\Delta E_{n p}^{\rm str} - \frac{i}{2} \Gamma_{n p}^{\rm str} = \frac{2 ( n^2 - 1 )}{n^5 \mu_c a_B^5} \, v_p + \cdots.
\label{sec4e}
\end{equation}
The corresponding coefficients are $1.875 \times 10^{-3}$ and $6.583 \times 10^{-4} \ {\rm eV / fm^3}$ for the 2$p$ and 3$p$ states, respectively. The physical $p$-wave amplitude determines the corresponding quantitative shift and width.

The $p$-wave sector has also been discussed phenomenologically. In an analysis of $e^+ e^- \to K^+ K^-$ data, \cite{PL2020} has interpreted a subthreshold pole introduced in fitting the CMD-3 data as a possible 2$p$ kaonium state, while the BABAR data were found to support an important strong-interaction contribution to its binding. The present calculation complements this picture by providing the low-lying $s$-wave spectrum and the electromagnetic $2 p \to 1 s$ transition scale.

Using Coulomb wave functions, the standard electric dipole transition rate \cite{HAB1957} gives
\begin{equation}
\Gamma_{2 p \to 1 s}^{E 1} = \frac{2^8}{3^8} \, \mu_c \alpha^5 = 1.99 \times 10^{-4} \ {\rm eV},
\label{secleq}
\end{equation}
corresponding to a radiative lifetime
\begin{equation}
\tau_{2 p \to 1 s} = 3.30 \times 10^{-12} \ {\rm s}.
\label{secfew}
\end{equation}
The point-Coulomb 2$p$ energy and the strong-shifted Fit 2 1$s$ energy give a transition energy of 4.996 keV, assuming the strong shift of the 2$p$ level is negligible. Including the leading additive vacuum polarization and finite-size corrections raises this to approximately 5.009 keV. Overall, the transition linewidth is dominated by the 1$s$ width and is approximately 1.67 keV.

\subsection{Fit 2 parameter dependence and systematic variations}
\label{IVD}

\begin{figure*}[t]
\centering
\includegraphics[width = \textwidth]{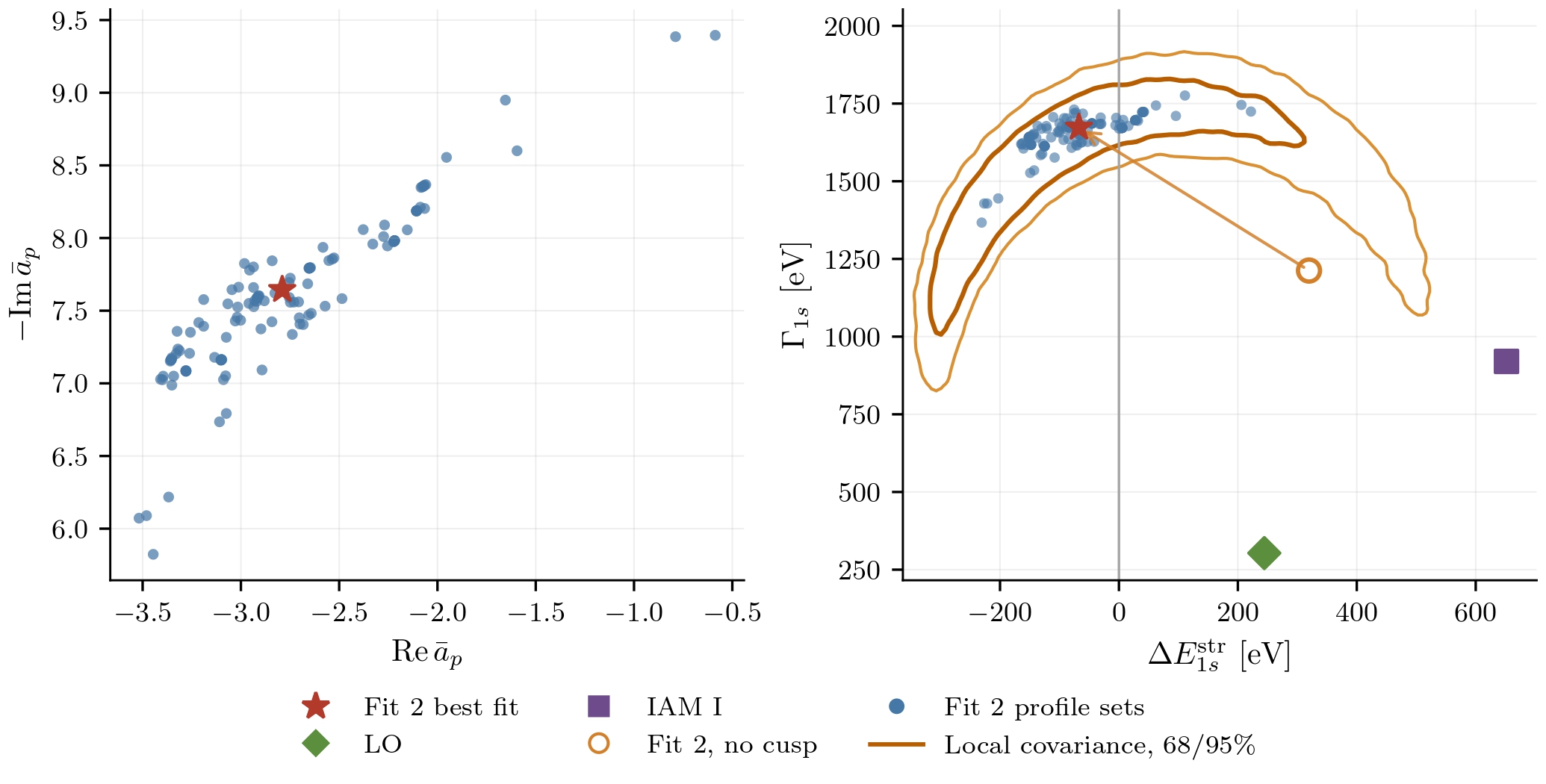}
\caption{Fit 2 parameter dependence of the charged scattering length and the kaonium ground-state observables. The left panel shows the Fit 2 profile sets, and the right panel shows the corresponding strong shifts and widths, along with the 68\% and 95\% local covariance contours. The no-cusp point and the effect of the neutral kaon cusp are also shown. The LO and IAM I results are included for comparison.}
\label{fig3n}
\end{figure*}

We repeat the calculation for the Fit 2 best fit and profile sets from the public source \cite{FKG}. These profile sets satisfy $\Delta \chi^2 \simeq 1$ relative to the fit minimum. We obtain:
\begin{equation}
\begin{aligned}
-231 &\leq \Delta E_{1 s}^{\rm str} \leq 222~{\rm eV}, \\
1367 &\leq \Gamma_{1 s} \leq 1776~{\rm eV}.
\end{aligned}
\label{e1d}
\end{equation}
We also generate $50 \, 000$ multivariate-normal parameter draws using the local covariance from the Fit 2 result obtained with Minuit's MIGRAD minimizer in the public notebook \cite{FKG}, where the chiral low-energy constant $L_6$ is kept fixed. The 16th--84th percentile ranges are then
\begin{equation}
\begin{aligned}
-245 &\leq \Delta E_{1 s}^{\rm str} \leq 237~{\rm eV}, \\
1316 &\leq \Gamma_{1 s} \leq 1740~{\rm eV}.
\end{aligned}
\label{e2d}
\end{equation}
The width distribution is asymmetric. Its median is 1594 eV, while the best fit value is 1673.4 eV. The full percentiles are given in Table~\ref{tab4d}.
\begin{table}[t]
\caption{Uncertainty propagation for Fit 2 using the local covariance matrix. The entries are percentiles of $50 \, 000$ multivariate-normal parameter draws. The energy shift and width are given in eV.}
\centering \small
\begin{tabular}{@{}lrrrrr@{}}
\toprule
Quantity & 2.5\% & 16\% & 50\% & 84\% & 97.5\% \\
\midrule
$\Delta E^{\rm str}_{1 s}$ & $-307$ & $-245$ & $-68$ & 237 & 467 \\
$\Gamma_{1 s}$ & 1025 & 1316 & 1594 & 1740 & 1839 \\
\bottomrule
\end{tabular}
\label{tab4d}
\end{table}
We have also shown the Fit 2 profile sets in the charged scattering length plane and the local covariance contours in the shift--width plane in Fig.~\ref{fig3n}.
The neutral kaon cusp has a sizable effect. Without the cusp correction, the two Fit 2 isospin amplitudes give approximately $( \Delta E_{1 s}^{\rm str}, \Gamma_{1 s} ) = (320, 1213)~{\rm eV}$. Including the cusp lowers the shift by about 387 eV and increases the width by about 461 eV.

\begin{figure}[t]
\centering
\includegraphics[width = 0.49\textwidth]{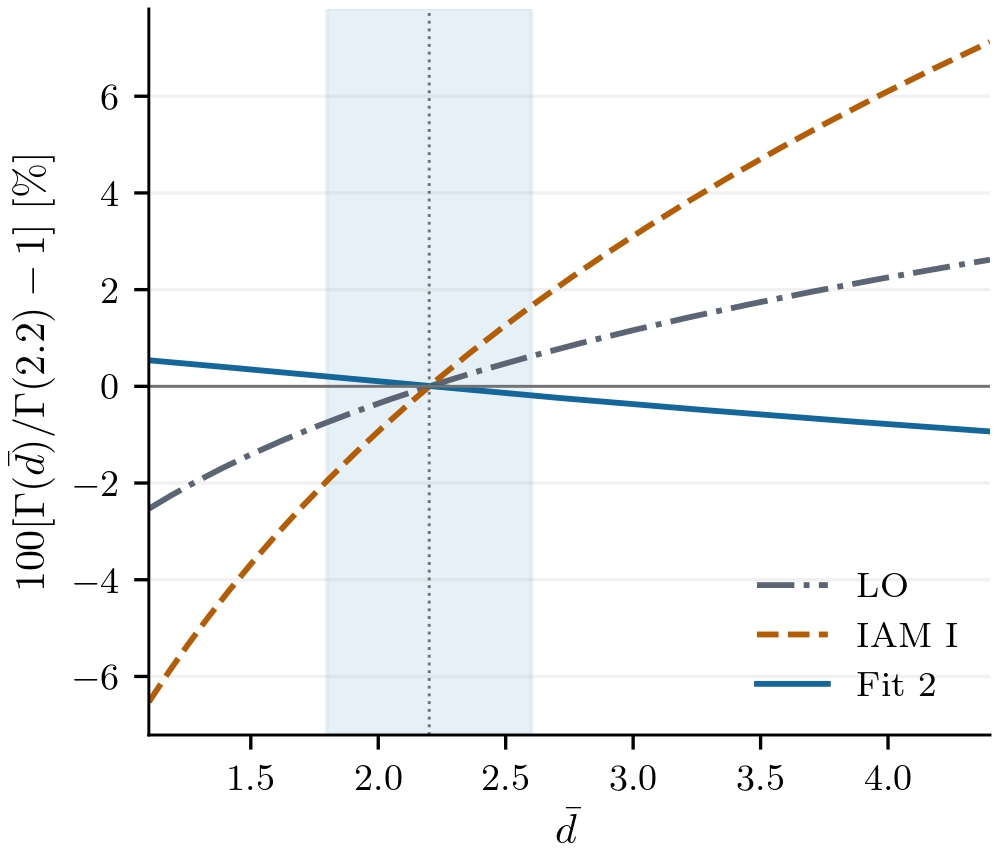}
\caption{Relative change in the kaonium ground-state width with the matching radius $\bar{d}$, using $\bar{d} = 2.2$ as the reference value. The shaded region shows $1.8 \leq \bar{d} \leq 2.6$. Results are shown for LO, IAM I, and Fit 2.}
\label{fig10}
\end{figure}

We then vary the matching radius $\bar{d}$. For $1.8 \leq \bar{d} \leq 2.6$, the IAM I width varies from 902.3 to 935.5 eV, while the Fit 2 width varies from 1670.3 to 1676.7 eV. For the wider interval $1.1 \leq \bar{d} \leq 4.4$, the ranges become $[860.2, 985.8]$ eV for IAM I and $[1657.8, 1682.5]$ eV for Fit 2. Figure~\ref{fig10} shows the relative change in the width as $\bar{d}$ is varied, using $\bar{d} = 2.2$ as the reference value.

The remaining variations are collected in Table~\ref{tabng}.
\begin{table*}[t]
\caption{Fit 2 parameter ranges and other variations, in eV. The first two rows give absolute ranges of $(\Delta E_{1 s}^{\rm str}, \Gamma_{1 s})$; the remaining rows give changes from the best fit result $(-67.3, 1673.4)$~eV. The vacuum polarization plus finite size row gives an additive electromagnetic correction to the energy relative to the point-Coulomb level.}
\begin{ruledtabular}
\begin{tabular}{lcc}
Variation & Shift [eV] & Width [eV] \\
\hline
Local covariance, 16--84\% & $[-245, 237]$ & $[1316, 1740]$ \\
Fit 2 profile sets & $[-231, 222]$ & $[1367, 1776]$ \\
Omit neutral kaon cusp & $+386.9$ & $-460.8$ \\
Alternative matching prescription & $+74.2$ & $+11.2$ \\
$\bar{d} = 1.8$--2.6 & $[-8.9, +10.8]$ & $[-3.2, +3.3]$ \\
$\bar{d} = 1.1$--4.4 & $[-36.6, +37.5]$ & $[-15.6, +9.0]$ \\
Charged kaon threshold & $[-9.6, -9.3]$ & $[-23.7, -23.5]$ \\
Vacuum polarization + finite-size correction & $-12.85(5)$ & --- \\
Vacuum polarization correction to strong contribution & $[-2.0, -0.5]$ & $[+11.0, +12.6]$ \\
Effective range & $-0.20$ & $-0.11$
\end{tabular}
\end{ruledtabular}
\label{tabng}
\end{table*}
Evaluating the strong amplitude directly at the physical charged kaon threshold changes $(\Delta E_{1 s}^{\rm str}, \Gamma_{1 s})$ by about $(-9.3, -23.5)~{\rm eV}$, while the effective range correction is only $(-0.20, -0.11)~{\rm eV}$. The alternative matching prescription, defined in Sec.~\ref{IIIA}, changes the shift and width by $(+74.2, +11.2)$ eV. The leading vacuum polarization and finite-size corrections shift the level by $-12.85(5)$ eV. The vacuum polarization correction to the strong contribution is listed separately.

\section{Photon-fusion line shape and lifetime}
\label{V}

\subsection{Pole width and real-axis FWHM}
\label{VA}

The pole width and the full width at half maximum (FWHM) of a real-axis line shape are not generally the same. With the pole energy measured relative to the charged kaon threshold, we define
\begin{equation}
B_{1 s} = -\operatorname{Re} E_{1 s}, \qquad \Gamma_{1 s}^{\rm pole} = -2 \operatorname{Im} E_{1 s}.
\end{equation}
For an isolated pole with constant residue, the line shape is Lorentzian and its FWHM is equal to the pole width. For a more general line shape, the energy dependence of the amplitude can change the FWHM, so the two widths may differ.

\begin{figure*}[t]
\centering
\includegraphics[width = \textwidth]{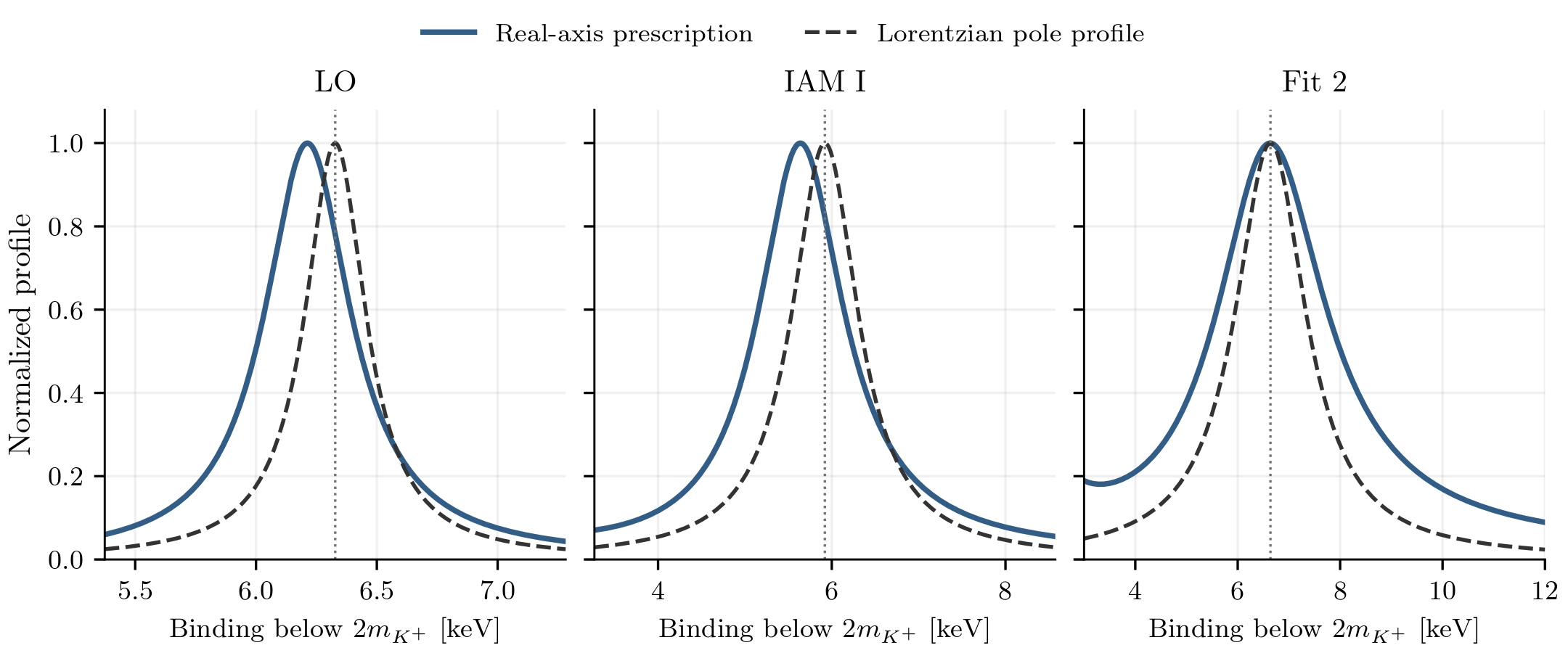}
\caption{Comparison of the real-axis prescription of \cite{AB2026} (solid curves) with Lorentzian profiles constructed from the KP poles (dashed curves), using the LO, IAM I, and Fit 2 inputs, and the physical charged kaon mass. Each curve is normalized to its own maximum. The dotted lines mark the KP pole binding energies.}
\label{figi4}
\end{figure*}

For the LO strong-interaction input, the KP equation gives
\begin{equation}
B_{1 s} = 6.3278 \ {\rm keV}, \qquad \Gamma_{1 s}^{\rm pole} = 303.3 \ {\rm eV}.
\end{equation}
We compare the real-axis prescription of \cite{AB2026} with Lorentzian profiles constructed from the KP poles, using the physical charged kaon mass and the same strong input. For LO, the prescription of \cite{AB2026} gives a peak binding energy of 6.2126 keV and a FWHM of 433.5 eV. For IAM I, the corresponding values are 5.6383 keV and 1.2062 keV. For Fit 2, they are 6.6338 keV and 2.6486 keV, compared with a KP pole binding energy of 6.6395 keV and a pole width of 1.6734 keV. Figure~\ref{figi4} compares these line shapes with the corresponding pole profiles.

We next examine how the larger kaonium width affects the photon-fusion line shape after experimental smearing.

\subsection{Photon-fusion line shape}
\label{VB}

\begin{figure*}[t]
\centering
\includegraphics[width = \textwidth]{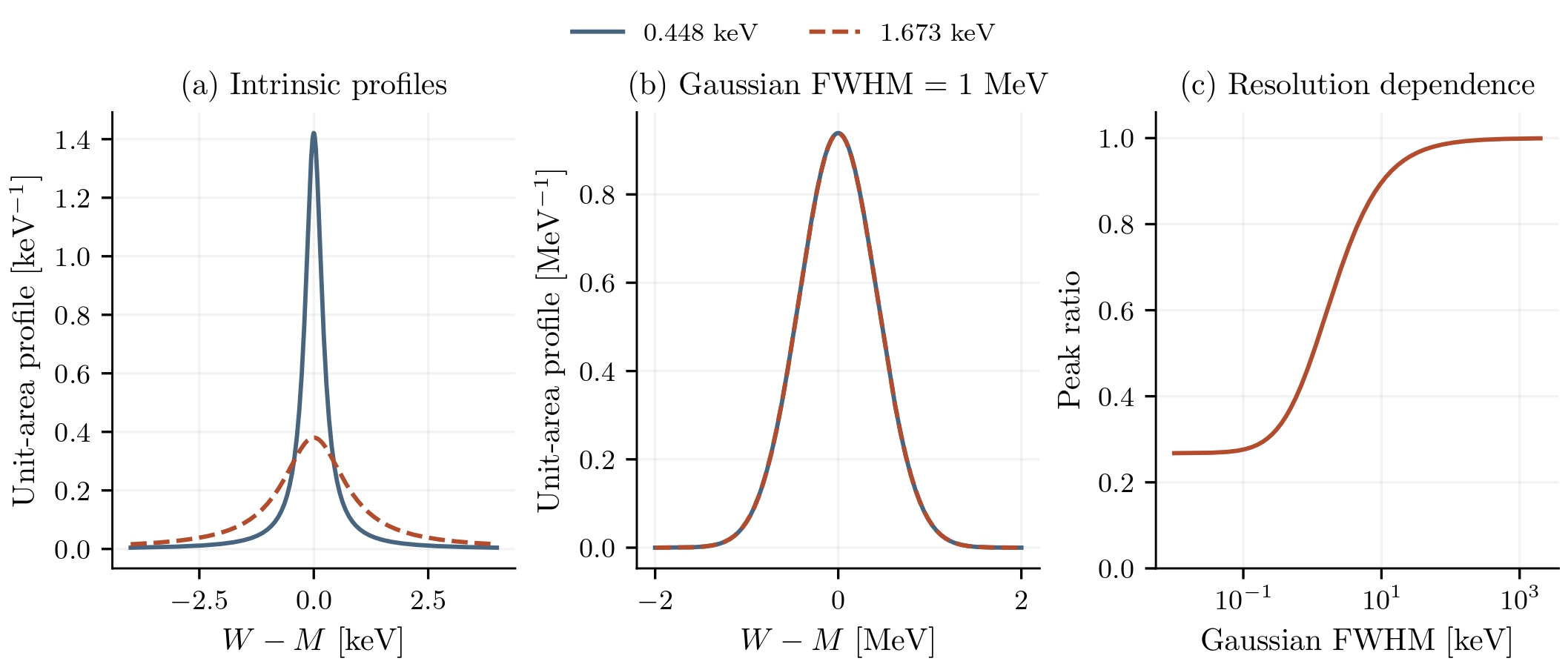}
\caption{Effect of the kaonium width on the photon-fusion pole profile. Panel (a) shows unit-area Lorentzian profiles with widths 0.448 and 1.673 keV at a common resonance position. Panel (b) shows the two profiles after smearing with a Gaussian response with a FWHM of 1 MeV. Panel (c) shows how the ratio of their peak heights changes with the Gaussian FWHM, taking the 1.673 keV profile relative to the 0.448 keV profile. The pole area is kept fixed throughout.}
\label{figi5}
\end{figure*}

Photon-fusion production of kaonium has been discussed in \cite{DE2025, AB2026}. Here we consider the isolated kaonium contribution to
\begin{equation}
\gamma \gamma \to f, \qquad f = \pi^0 \pi^0, \ \pi^0 \eta.
\end{equation}
We write $M = 2 m_{K^+} - B_{1 s}$, assume $\Gamma \simeq \Gamma_{1 s}^{\rm pole}$, and neglect electromagnetic decays in the total width. For unpolarized photons and a scalar resonance, the pole contribution can be written in the Breit--Wigner form
\begin{equation}
\sigma_f^{\rm pole} ( W ) \simeq \frac{8 \pi \, \Gamma_{\gamma \gamma} \Gamma_f}{( W^2 - M^2 )^2 + M^2 \Gamma^2}.
\label{1bw}
\end{equation}
Near the resonance, with $\epsilon = W - M$ and $\mathcal{B}_f = \Gamma_f / \Gamma$, (\ref{1bw}) becomes
\begin{equation}
\sigma_f^{\rm pole} ( W ) \simeq \frac{2 \pi}{M^2} \frac{\Gamma_{\gamma \gamma} \Gamma \mathcal{B}_f}{\epsilon^2 + \Gamma^2 / 4}.
\label{rbw}
\end{equation}
The peak value and integrated pole contribution are then
\begin{equation}
\begin{aligned}
\sigma_f^{\rm pole} ( M ) &= \frac{8 \pi}{M^2} \frac{\Gamma_{\gamma \gamma} \mathcal{B}_f}{\Gamma}, \\
\int dW \, \sigma_f^{\rm pole} ( W ) &\simeq \frac{4 \pi^2}{M^2} \Gamma_{\gamma \gamma} \mathcal{B}_f.
\end{aligned}
\label{3fisr}
\end{equation}
Thus, at fixed $\Gamma_{\gamma \gamma} \mathcal{B}_f$, increasing the width lowers the unsmeared peak but leaves the integrated pole contribution unchanged. Reference~\cite{AB2026} has used a Breit--Wigner width of 0.448 keV. For comparison with the Fit 2 result, we use widths of 0.448 and 1.673 keV at a common resonance position. After convolution with a Gaussian response of FWHM 1 MeV, the observed FWHM changes only from 1.0002 to 1.0009 MeV. At fixed integrated pole contribution, the peak decreases by only 0.115\%.
Figure~\ref{figi5} shows this comparison.

The width effect is also very small for Belle's experimental resolution in $\gamma \gamma \to \pi^0 \eta$ \cite{SU2009}. Using Belle's parameterization \cite{SU2009} and defining $u = W / ( 1 \ {\rm GeV} )$, the standard deviations on the low- and high-mass sides are
\begin{equation}
\begin{aligned}
\sigma_W^{<} &= 0.013 u \left( 1.4 - \frac{0.3}{u^2} \right) {\rm GeV}, \\
\sigma_W^{>} &= 0.0077 u \left( 1.4 - \frac{0.3}{u^2} \right) {\rm GeV}.
\end{aligned}
\label{stdb}
\end{equation}
At the present pole mass, these are about 14.0 and 8.30 MeV. We model the detector response with a normalized split Gaussian using the two standard deviations in (\ref{stdb}). It turns out that increasing the width from 0.448 to 1.673 keV, while the resonance position and pole area are kept fixed, changes the contribution to the 980--1000 MeV bin by only 0.0036\%. Changing the resonance position from 991.994 MeV in \cite{AB2026} to 987.347 MeV here, together with the width change, reduces the contribution to the same bin by about 8\%. Note that the mass and binding energy quoted in \cite{AB2026} imply a threshold of approximately 992 MeV, while the present calculation uses the physical charged kaon threshold $2 m_{K^+} = 987.354$ MeV.

\subsection{Lifetime and propagation in matter}
\label{VC}

The Fit 2 ground-state pole width gives
\begin{equation}
\tau_{1 s} = \frac{\hbar}{\Gamma_{1 s}^{\rm pole}} = 3.93 \times 10^{-19} \ {\rm s}, \quad c \tau_{1 s} = 0.118 \ {\rm nm}.
\label{e1vc}
\end{equation}
For comparison, the measured pionium ground-state lifetime is approximately $3.15 \times 10^{-15}$ s, about $8 \times 10^3$ times longer. DIRAC has also observed long-lived excited pionium states with lifetimes of order $10^{-11}$ s and longer \cite{BA2015}.

The competition between break-up and annihilation has been studied for pionium propagating through matter \cite{CS2003}. To estimate its effect for kaonium, we consider an atom moving at constant speed through a uniform target. We neglect transitions between bound states and treat break-up and decay as independent processes. Their rates in the target rest frame are
\begin{equation}
r_{\rm br} = n_T \sigma_{\rm br} \beta c, \qquad r_{\rm dec} = \frac{1}{\gamma \tau},
\label{smf}
\end{equation}
where $n_T$ is the target number density, $\sigma_{\rm br}$ is the break-up cross section, $\tau$ is the proper lifetime, $\beta c$ is the atomic speed, and $\gamma$ is the Lorentz factor. For an infinite target, the probability of break-up before decay is
\begin{equation}
\begin{aligned}
P_{\rm br}^{(\infty)} &= \int_0^\infty dt \, r_{\rm br} e^{-( r_{\rm br} + r_{\rm dec} ) t} \\
&= \frac{r_{\rm br}}{r_{\rm br} + r_{\rm dec}} = \frac{\zeta}{1 + \zeta},
\end{aligned}
\label{fevc2}
\end{equation}
with $\zeta = r_{\rm br} / r_{\rm dec} = n_T \sigma_{\rm br} \beta \gamma c \tau$. For $n_T = 10^{23} \ {\rm cm}^{-3}$ and $\beta \gamma = 10$, a ground-state break-up probability of 10\% requires $\sigma_{\rm br} \simeq 9.4 \times 10^{-18} \ {\rm cm}^2$. To compare with pionium, by assuming $\sigma_{\rm br} \propto a_B^2$, we obtain a ground-state kaonium break-up-to-decay rate ratio about $10^{-5}$ times the pionium value at the same boost $\beta \gamma$ and target density.

Using the widths in Table~\ref{tabiii}, it can be shown that the 2$s$ and 3$s$ lifetimes exceed the ground-state lifetime by factors of 8.09 and 27.43, respectively. States with nonzero orbital angular momentum may live much longer because the leading short-range $s$-wave annihilation term is absent. Using Coulomb wave functions, the electric dipole rate evaluated in Sec.~\ref{IVC} gives the partial radiative lifetime
\begin{equation}
\tau_{2 p \to 1 s} = 3.30 \times 10^{-12} \ {\rm s},
\end{equation}
consistent with the earlier estimate in \cite{PL2020}. Note that strong $p$-wave annihilation and other decay channels can shorten the total 2$p$ lifetime.

\section{Conclusions}
\label{VI}

Using Fit 2 and IAM I as two fitted coupled-channel one-loop chiral descriptions of the near-threshold $K \bar{K}$ interaction, we obtain ground-state kaonium decay widths of 1.67 and 0.92 keV, respectively. The comparison of the two widths shows that the kaonium width depends not only on the absorptive part of the $K^+ K^-$ amplitude but also on its real part through the Coulomb correction.

Despite the increased kaonium width, particularly for Fit 2, the first three $s$-wave levels remain spectrally distinct, with the 2$s$ and 3$s$ states narrower than the ground state. At fixed pole area and resonance position, the increased intrinsic width has a negligible effect on the photon-fusion line shape after including Belle's experimental resolution. We have also examined states with nonzero orbital angular momentum and calculated the $2 p \to 1 s$ radiative transition rate.

Further improvement of the kaonium prediction would benefit from dispersive or other data-driven determinations of, in particular, the complex $I = 0$ and $I = 1$ $K \bar{K}$ threshold amplitudes. A more complete treatment of electromagnetic and isospin-breaking corrections would further improve the prediction of the atomic observables. Calculations of the $p$-wave threshold amplitude, together with kaonium formation probabilities and cascade rates, could also help assess the experimental prospects for the excited states of this elusive but fascinating atom.

\appendix

\section{Numerical implementation and local dependence of the width}
\label{app}

\subsection{Numerical implementation}

Direct evaluation of the closed-form scalar loop functions used in the one-loop amplitudes \cite{AGN2002} can lose numerical precision near $s = 0$. We therefore evaluate the subtracted loop in this region using the equivalent Feynman parameter representation:
\begin{equation}
\bar{J} ( s; m_1, m_2 ) = -\frac{1}{16 \pi^2} \int_0^1 dx \, \ln \left[ 1 - s \, w(x) \right],
\end{equation}
where
\begin{equation}
w(x) = \frac{x ( 1 - x )}{x m_1^2 + ( 1 - x ) m_2^2}.
\end{equation}
For sufficiently small $|s|$, this gives
\begin{equation}
\bar{J} (s) = \sum_{j = 1}^{\infty} c_j s^j, \qquad c_j = \frac{1}{16 \pi^2 j} \int_0^1 dx \, w(x)^j.
\end{equation}
The twice subtracted loop, $\bar{J}(s) - s \bar{J}'(0)$, is obtained by starting the series at $j = 2$. For Fit 2, we evaluate the partial-wave projection numerically and extract the threshold parameters by fitting the effective range expansion near threshold. At the best fit point, varying the angular quadrature and fitting interval changes both the kaonium shift and width by less than $5 \times 10^{-4}$ eV.

\subsection{Local dependence of the width}

The local dependence relation (\ref{feis}), in the main text, follows from the zero range KP equation at fixed $\bar{d} = 2.2$, with $\bar{a}_d = \bar{a}_p$. For
\begin{equation}
F(\lambda) = \frac{\lambda}{2} + \ln \lambda + \psi \left( 1 - \frac{1}{\lambda} \right) + \gamma_E,
\end{equation}
implicit differentiation gives
\begin{equation}
\frac{\partial E_{1 s}}{\partial \bar{a}_p} = \frac{m_{K^+} \alpha \lambda_1}{2 \bar{a}_p^2 F'(\lambda_1)},
\end{equation}
where $\lambda_1$ is the ground-state KP root and
\begin{equation}
F'(\lambda) = \frac{1}{2} + \frac{1}{\lambda} + \frac{\psi^{(1)} ( 1 - 1 / \lambda )}{\lambda^2},
\end{equation}
with $\psi^{(1)}$ as the trigamma function. At the Fit 2 best fit point,
\begin{equation}
\frac{\partial E_{1 s}}{\partial \bar{a}_p} = ( 96.66 + 57.24 i ) \ {\rm eV}.
\end{equation}
To write the response in terms of the isospin scattering lengths, we define
\begin{equation}
\rho = \frac{m_{K^+}}{m_K}, \qquad q_0 = \frac{\kappa_0}{m_K}, \qquad h = 1 - \frac{q_0}{2} ( \bar{a}_0 + \bar{a}_1 ).
\end{equation}
Differentiating the neutral kaon cusp relation gives
\begin{equation}
\frac{\partial \bar{a}_p}{\partial \bar{a}_0} = \frac{\rho ( 1 - q_0 \bar{a}_1 )^2}{2 h^2}, \qquad \frac{\partial \bar{a}_p}{\partial \bar{a}_1} = \frac{\rho ( 1 - q_0 \bar{a}_0 )^2}{2 h^2}.
\end{equation}
Using $\Gamma_{1 s} = -2 \, \operatorname{Im} E_{1 s}$, the first-order variation is
\begin{equation}
\delta \Gamma_{1 s} \simeq -2 \, \operatorname{Im} \! \left[ \frac{\partial E_{1 s}}{\partial \bar{a}_p} \left( \frac{\partial \bar{a}_p}{\partial \bar{a}_0} \, \delta \bar{a}_0 + \frac{\partial \bar{a}_p}{\partial \bar{a}_1} \, \delta \bar{a}_1 \right) \right].
\end{equation}
Evaluating these derivatives at the Fit 2 best fit point gives
\begin{equation*}
\begin{aligned}
\frac{\delta \Gamma_{1 s}}{{\rm eV}} \simeq{} &80.4 \, \delta \operatorname{Re} \bar{a}_0 - 60.4 \, \delta \operatorname{Im} \bar{a}_0 \\
&-230.3 \, \delta \operatorname{Re} \bar{a}_1 - 29.3 \, \delta \operatorname{Im} \bar{a}_1,
\end{aligned}
\end{equation*}
as quoted in (\ref{feis}) of the main text.

\end{document}